\documentclass[a4paper,11pt]{article}
\pdfoutput=1 

\usepackage{jcappub} 

\usepackage[T1]{fontenc} 

\newcommand{\nnine}{\texttt{n9}}
\newcommand{\treco}{$t_{reco}$}
\newcommand{\tmc}{$t_{MC}$}

\title{Supernova model discrimination with a kilotonne-scale Gd-H$_{2}$O Cherenkov detector}

\author[a,1]{Y. Schnellbach,\note{Corresponding author.}}
\author[b]{J. Migenda,}
\author[a]{A. Carroll,}
\author[a]{J. Coleman,}
\author[b]{L. Kneale,}
\author[b]{M. Malek,}
\author[a]{C. Metelko}
\author[a]{and A. Tarrant}

\affiliation[a]{Department of Physics, University of Liverpool, Oliver Lodge Laboratory, Oxford Street, Liverpool, L69 7ZE, U.K.}
\affiliation[b]{Department of Physics and Astronomy, University of Sheffield, Hicks Building, Hounsfield Road, Sheffield, S3 7RH, U.K.}

\emailAdd{y.schnellbach@liverpool.ac.uk}
\emailAdd{jost.migenda@kcl.ac.uk}
\emailAdd{a.e.carroll@liverpool.ac.uk}
\emailAdd{j.coleman@liverpool.ac.uk}
\emailAdd{e.kneale@sheffield.ac.uk}
\emailAdd{m.malek@sheffield.ac.uk}
\emailAdd{carl.metelko@liverpool.ac.uk}
\emailAdd{a.tarrant@liverpool.ac.uk}

\abstract{The supernova model discrimination capabilities of the WATCHMAN detector concept are explored. This cylindrical kilotonne-scale water Cherenkov detector design has been developed to detect reactor antineutrinos through inverse $\beta$-decay for non-proliferation applications but also has the ability to observe antineutrino bursts of core-collapse supernovae within our galaxy. Detector configurations with sizes ranging from 16\,m to 22\,m tank diameter and 10\% to 20\% PMT coverage are used to compare the expected observable antineutrino spectra based on the Nakazato, Vartanyan and Warren supernova models. These spectra are then compared to each other with a fixed event count of 100 observed inverse $\beta$-decay events and a benchmark supernova at 10\,kpc distance from Earth. By comparing the expected spectra, each detector configuration's ability to distinguish is evaluated. This analysis then demonstrates that the detector design is capable of meaningful event discrimination (90+\% accuracy) with 100 observed supernova antineutrino events in most configurations. Furthermore, a larger tank configuration can maintain this performance at 10\,kpc distance and above, indicating that overall target mass is the main factor for such a detector's discrimination capabilities. Finally, it is estimated that the detector design can provide early warning capability for supernova bursts for the entire Milky Way in all configurations.}


\begin{document}
\maketitle
\flushbottom

\section{Introduction}
\label{sec:intro}

The WATCHMAN Collaboration (WATer CHerenkov Monitor for ANtineutrinos) has developed the WATCHMAN detector concept~\cite{burns2018remote, osti_1544490} for the Neutrino Experiment One (NEO) at the planned Advanced Instrumentation Testbed (AIT) facility. The chief goal of AIT-NEO and the WATCHMAN Collaboration is the development of a detector capable of verifying the presence or absence of an active nuclear reactor through its antineutrino signature in the mid- to far-field, i.e. at a range of 10s to 100s of kilometres. Such capability is especially useful for reactor monitoring in a non-proliferation context~\cite{RevModPhys.92.011003}. However, a detector capable of monitoring reactor antineutrinos is also capable of observing astrophysical sources of antineutrinos, opening the door to collaborative and scientific uses of such a detector.

Core-collapse supernovae are rare but extremely energetic events in the universe, providing a high flux of observable antineutrinos~\cite{Adams_2013}. As neutrinos and antineutrinos emitted during the core-collapse carry information from the core of the progenitor star, unobstructed by the outer layers~\cite{JANKA200738}, energy and time measurements of the antineutrinos allow to test existing supernova models and guide the development of future models. Due to the limited number of supernova antineutrinos observed during the SN1987 supernova~\cite{Vissani_2014}, current developments are limited by the sparse data set. Furthermore, since neutrino signals from supernovae are expected before the optical signal, they are also a key piece in modern multi-messenger astronomy~\cite{Snews, Al_Kharusi_2021} and the expected signal rates determine a detector's effective range as an early warning station. 

To understand the physics potential of the WATCHMAN detector concept, its ability to distinguish different supernova models based on the detectable spectrum is studied. This discrimination power serves as a benchmark for the detector's ability to extract relevant information from an observed supernova. Similar studies using the same underlying methodology have been conducted for large scale neutrino detectors~\cite{HyperKDiscrimination}, but due to the much smaller scale of the WATCHMAN detector concept, it is important to understand how design choices, such as detector size and photosensor coverage, impact the physics potential.

\section{Antineutrino detection in WATCHMAN}
\label{sec:watchman}

\begin{figure}
    \centering
    \includegraphics[width=0.5\textwidth]{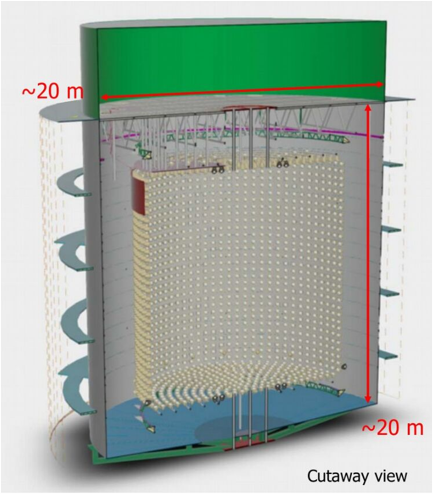}
    \caption{Schematic cutaway view of the WATCHMAN detector concept by Jan Boissevain (University of Pennsylvania), showing the outer tank wall, the array of PMTs and the inner volume enclosed by the PMTs support structure.}
    \label{fig:watchman}
\end{figure}

The WATCHMAN detector concept consists of a kilotonne-scale upright cylindrical tank, shown in figure~\ref{fig:watchman}, containing ultra-pure water doped with 0.2\% gadolinium sulphate (Gd$_{2}$(SO$_{4}$)$_{3}$), yielding an effective Gd concentration of 0.1\%. Inside the tank is a support frame, holding an array of photomultiplier tubes (PMTs) as photosensors, forming an inner detector (ID) volume for detection and an outer volume acting as a buffer and veto volume. The detector configurations of interest are a smaller tank with 16\,m diameter and an inner detector radius of 5.7\,m and a larger tank with 22\,m diameter and an inner detector radius of 9.0\,m. The height of the tank and ID are equal to the diameter of the tank or ID respectively. Both designs are based on a baseline PMT coverage of 15\%, resulting in 2,500 PMTs for the smaller design and 4,600 PMTs for the larger design. As this study aims to understand the impact of the different detector parameters on its supernova physics potential, intermediate tank sizes between these two designs are considered and lower (10\%) and higher (20\%) PMT coverages are studied as well to allow for correct interpolation between these design parameters.

Supernova antineutrinos can interact with the detector medium via inverse $\beta$-decay (IBD), elastic scattering (ES) and charged-current interaction (CC) on nuclei. For a large water-filled detector the dominant detection channel for the electron antineutrinos is the IBD interaction, hence this study only considers the IBD channel. In this interaction, an incoming antineutrino above the threshold energy of 1.8\,MeV interacts with the quasi-free protons in the water~\cite{VogelAndBeacom, STRUMIA200342}, resulting in a outgoing positron and neutron in the detector medium:

\begin{equation}
    \overline{\nu_{e}}+p\to e^++n.
\end{equation}

The outgoing positron immediately produces Cherenkov light in the detection medium, yielding a prompt light signal detectable by the PMTs, with a total light yield dependent on the particle's energy. As positron energy is related to the energy of the incoming antineutrino~\cite{VogelAndBeacom}, this allows the positron energy and thus the energy of the incoming antineutrino to be estimated. The neutron thermalises within the medium by undergoing a random walk, scattering off atoms until it is captured by a nucleus. The capture nucleus is left in an excited state and de-excites by releasing $\gamma$-rays. Due to the high neutron-capture cross section of gadolinium, at 0.1\% Gd concentration, circa 91\% of all neutron captures in WATCHMAN are on gadolinium nuclei with an average capture time of about 30\,$\mu$s ~\cite{2004gadzooks}. The excited gadolinium nucleus releases an 8\,MeV $\gamma$-ray cascade, with the energy spread distributed across multiple $\gamma$-rays. The remaining 9\% of neutron captures occur on hydrogen nuclei, resulting in the emission of a single 2.2\,MeV  $\gamma$-ray. The de-excitation $\gamma$-rays, whether from gadolinium or hydrogen, produce Compton electrons in the medium, which can be detected via their Cherenkov light emission. This provides a secondary, time-coincident signal in conjunction with the prompt signal of the positron, allowing for powerful background rejection of non-antineutrino signals within the detector.

The planned location for the AIT-NEO was the Boulby Underground Laboratory, operated by the Science and Technology Facilities Council (STFC). This laboratory is located in the rock salt layers of Boulby Mine in the North East of England, an active polyhalite mine operated by the ICL Group Ltd. The surrounding rock salt contributes to a low-radon environment, minimising background rates. Additionally, the laboratory is situated 1.1\,km underground, providing 2.8\,km water equivalent (km.w.e.) of cosmic ray shielding, attenuating the cosmic muon flux by approximately $\mathcal{O}(10^{6})$ compared to the surface~\cite{ROBINSON2003347}.

\section{Supernova event generation}
\label{sec:supernovasim}

\begin{figure}
    \centering
    \includegraphics[width=0.325\textwidth]{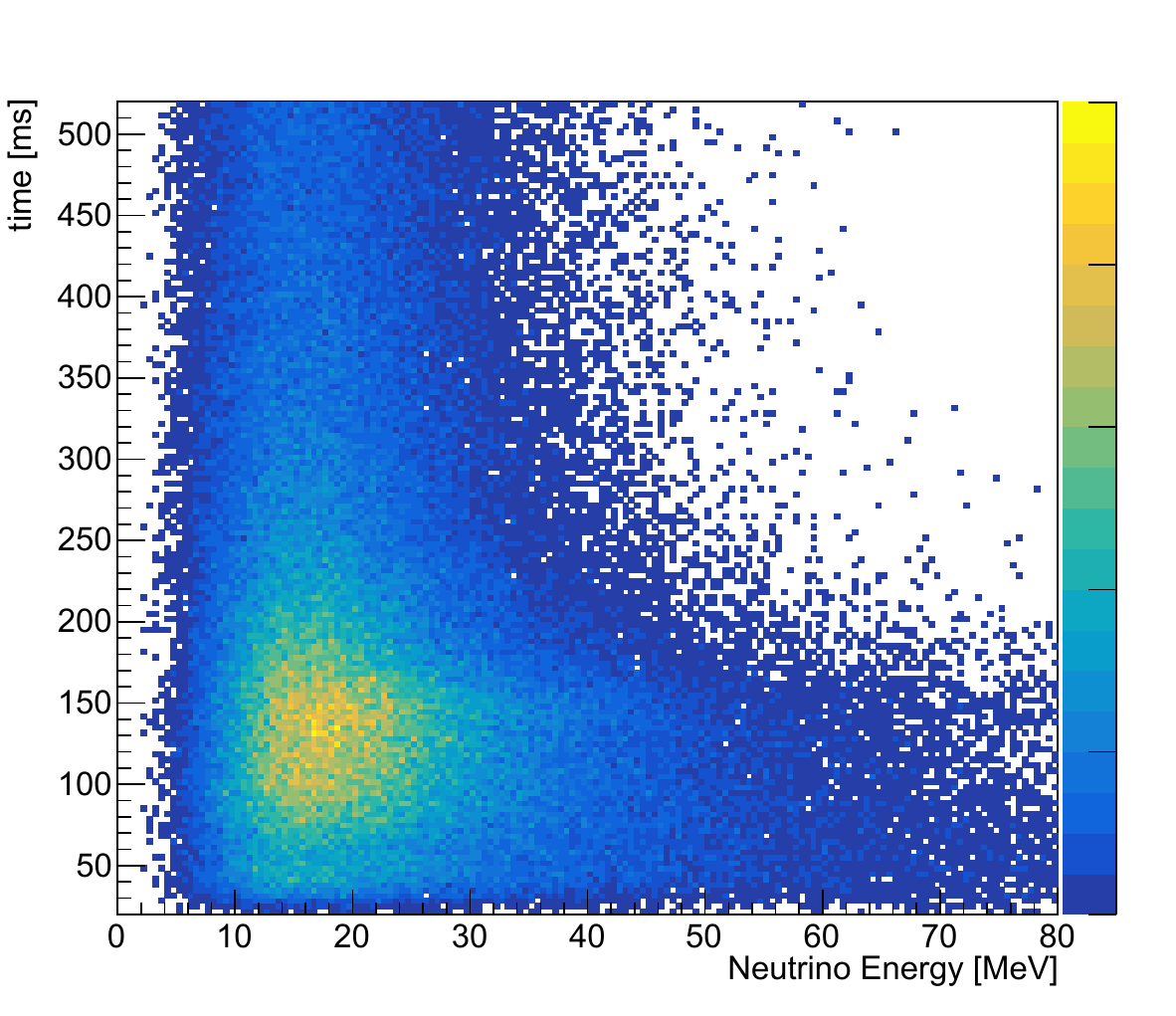}
    \includegraphics[width=0.325\textwidth]{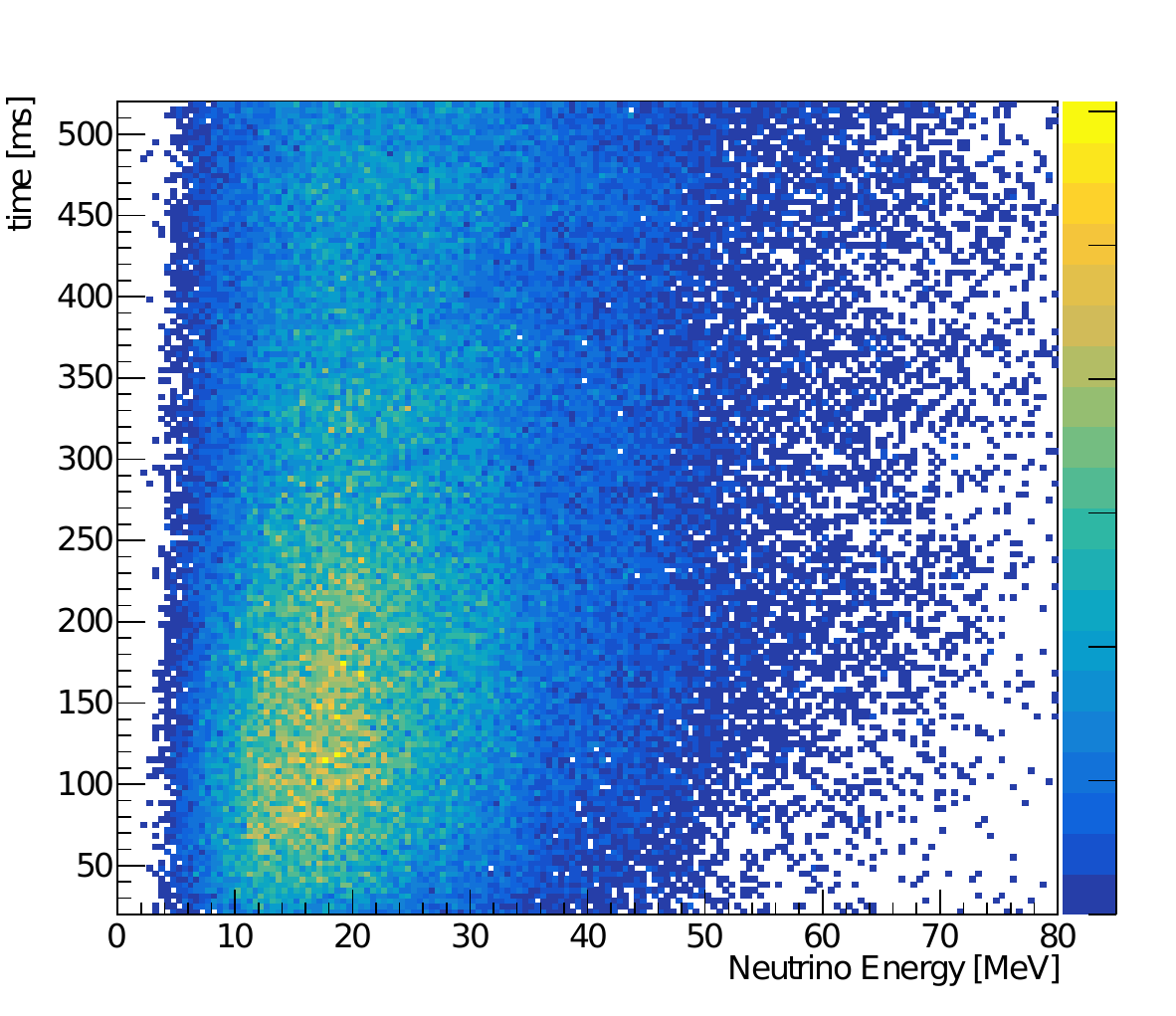}
    \includegraphics[width=0.325\textwidth]{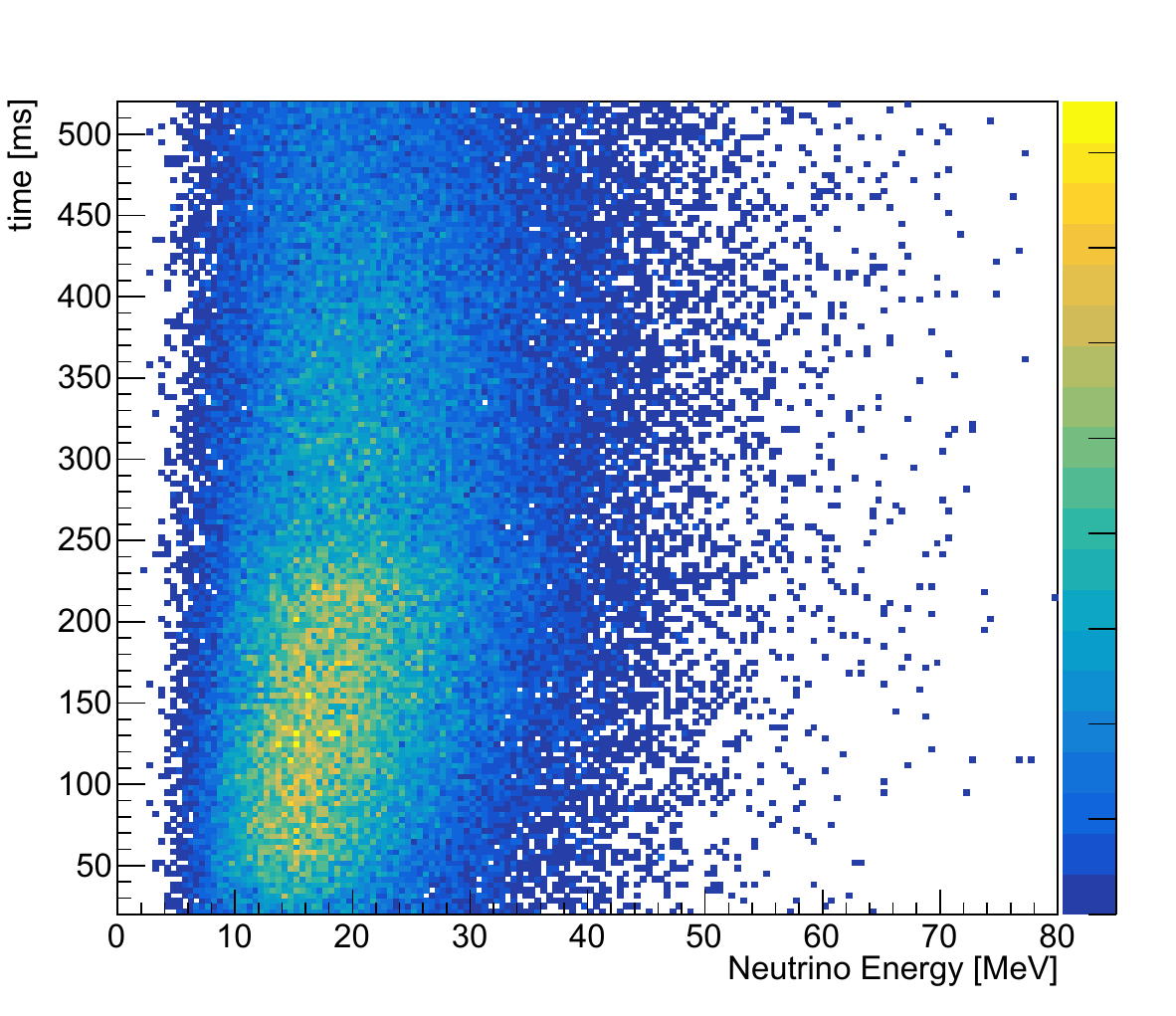}
    \caption{Observable electron antineutrino energies and time for the Nakazato (left), Vartanyan (center) and Warren (right) model. These spectra are convolved with the IBD cross-section, showing the observable spectrum within a detector.}
    \label{fig:modelspectra}
\end{figure}

The supernova IBD events within the detector are generated using \textsc{sntools}~\cite{Migenda2021}, a software package capable of taking high-fidelity model files and producing IBD events in the specified detector volume. For this study, three supernova models are considered:
\begin{itemize}
    \item The Nakazato family of models~\cite{Nakazato2013} provides simulations with different progenitor masses and metallicities. For this study, a 20\,$M_{\odot}$ progenitor with a metallicity of Z = 0.02 (solar metallicity) has been chosen. The Nakazato model is a widely-used model that provides a comparison point with other detectors.
    \item The Vartanyan model~\cite{Vartanyan2021} (via SNEWPY~\cite{Baxter:2021xyq,SNEWS:2021ewj}) uses two-dimensional FORNAX simulations to model core-collapse, providing a very different approach to supernova simulation than the Nakazato model. For this study, a 20\,$M_{\odot}$ progenitor star was chosen.
    \item The simulations for the Warren model~\cite{Warren2020} use a one-dimensional simulation approach called STIR (Supernova Turbulence In Reduced-dimensionality) to mimic the three-dimensional explosion mechanism. For this study, as for the Nakazato model, a 20\,$M_{\odot}$ progenitor star with a turbulence strength parameter $\alpha_\Lambda = 1.25$ was chosen. This model was chosen since it represents another recent approach to supernova modelling.
\end{itemize}
\textsc{sntools} is used with these models to generate events for a 500\,ms time interval from 20\,ms to 520\,ms post core bounce, containing the shock stagnation and accretion phase. This interval has been chosen because many models, including the Vartanyan and Warren model, only focus on the accretion phase. Furthermore, the earlier and later phases of a core-collapse supernova, neutronisation burst and proto-neutron star cooling respectively, are better understood and exhibit less model-dependent variations. Hence, they are of lesser interest to a model-discrimination study. The key variables of interest for a detector study are the electron antineutrino rate with respect to time and energy, as shown in figure~\ref{fig:modelspectra}, and energy and direction of the outgoing positron and neutron after the IBD interaction. \textsc{sntools} provides these values as an output file that can be read by a detector simulation to ascertain the subsequent detector response.

\section{Detector simulation and reconstruction}
\label{sec:simreco}

The interaction of the outgoing IBD positrons and neutrons with the detection medium and subsequent detection are handled by the RAT-PAC framework~\cite{ratpac}\footnote{The WATCHMAN collaboration employs a modified version with the code available at \texttt{https://github.com/AIT-WATCHMAN/rat-pac}}. This framework uses \textsc{Geant4}~\cite{AGOSTINELLI2003250, ALLISON2016186} for the underlying detector simulation and \textsc{Root}~\cite{fons_rademakers_2019_3457396} for data processing and analysis. RAT-PAC simulates the detector signal response on an event-by-event basis, including particle propagation through the detector medium, light emission and PMT response. The software also simulates the triggering and data acquisition of the detector, producing ROOT files containing the individual PMT hits with location and timing as well as information about the underlying event for subsequent analysis. The fidelity of the neutron capture simulation within RAT-PAC has also been improved using measurements from the DANCE calorimeter using a DICEBOX implementation~\cite{PhysRevC.84.014306} to model the capture on $^{157}$Gd, the gadolinium isotope with the highest capture cross section. Importantly, the RAT-PAC framework allows for dynamic adjustment of the detector parameters using a parametric construction. This capability is used to study the different detector configurations and their impact on the detector's discrimination ability.

The reconstruction uses an evolution of the BONSAI package, originally developed for Super-Kamiokande ~\cite{SMY2007118,CARMINATI2015666}, to determine the reconstructed energy and position of each particle. In order to minimise any background due to the radioactivity of the PMT glass, support structure or from background entering the detector from outside the tank, the detector volume of each studied detector configuration is fiducialised as listed in table~\ref{tab:detector}. This is implemented as a virtual fiducial volume, meaning that the reconstructed prompt vertex location is used to exclude any events reconstructed outside the volume of interest.

\begin{table}[ht]
    \centering
    \begin{tabular}{cccc}
    \hline
    Tank & Inner Detector & Fiducial & Fiducial\\
    Diameter [m]   & Radius [m] & Radius [m] & Mass [kt]\\
    \hline
    22 & 9.0 & 8.0 & 3.21\\
    18 & 7.7 & 6.7 & 1.89\\
    16 (a) & 6.7 & 5.7 & 1.16\\
    16 (b) & 5.7 & 4.7 & 0.65\\
    \hline
    \end{tabular}
    \caption{Overview of studied detector sizes, PMT positions defining the inner detector radius and the associated choice of the virtual fiducial volume and mass in the reconstruction and analysis. Tank height, inner detector height and fiducial height are equal to the respective diameters.}
    \label{tab:detector}
\end{table}

\begin{figure}
    \centering
    \includegraphics[width=0.49\textwidth]{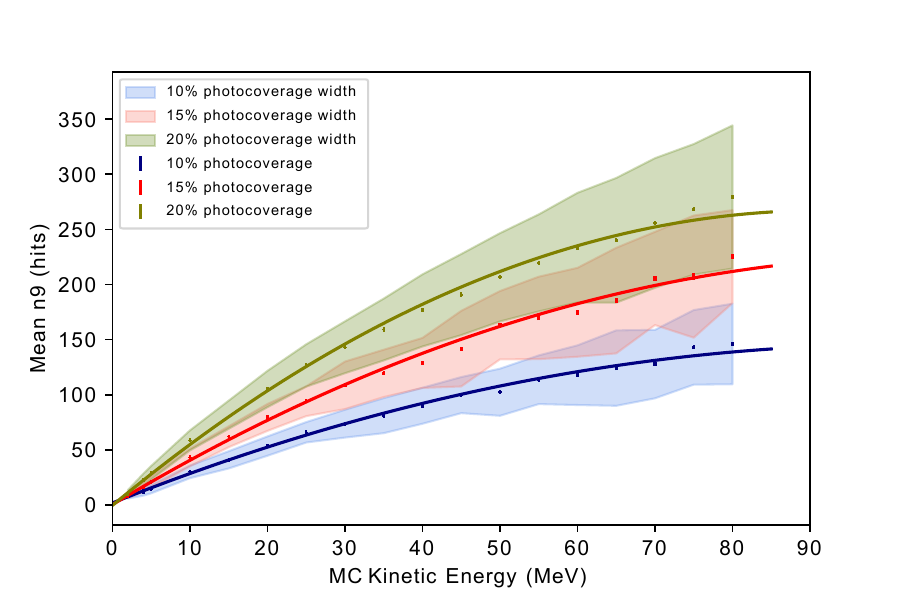}
    \includegraphics[width=0.49\textwidth]{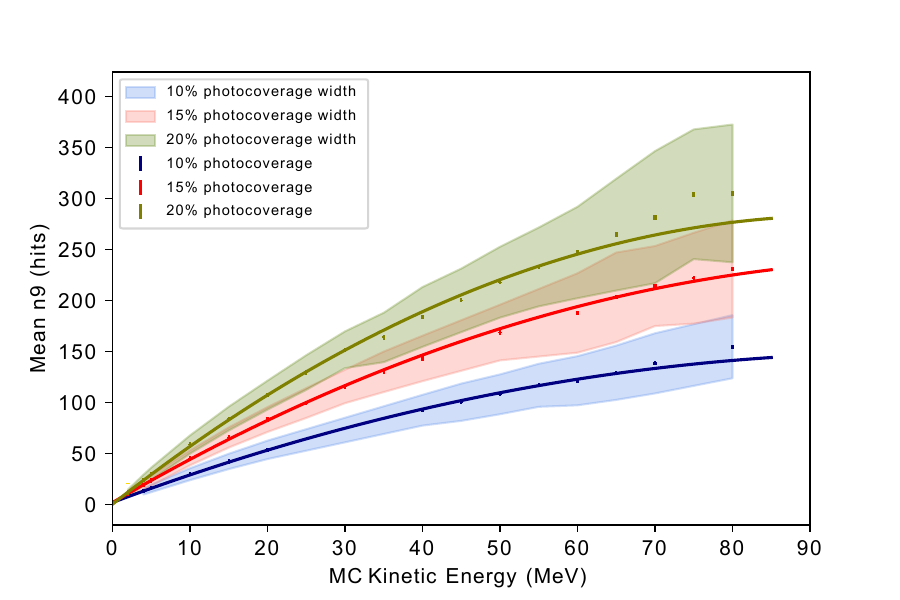}\\
    \includegraphics[width=0.49\textwidth]{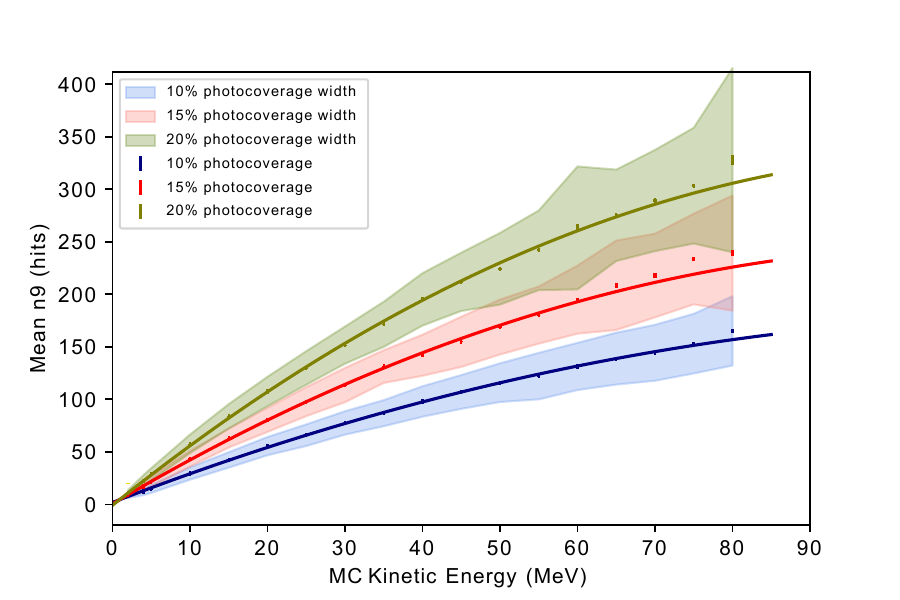}
    \includegraphics[width=0.49\textwidth]{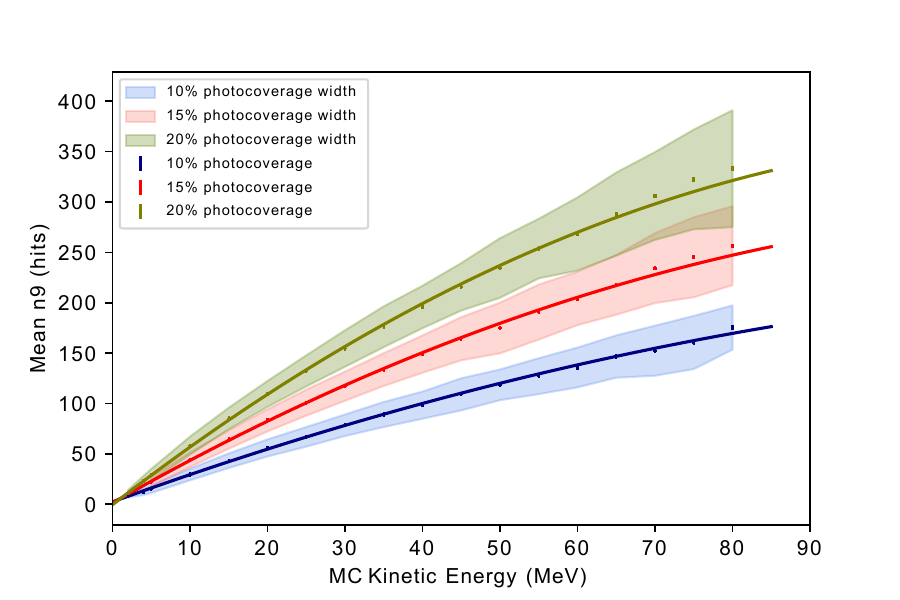}
    \caption{Mean energy spectra fitted for inner detector radii of 5.7\,m (top left), 6.7\,m (top right), 7.7\,m (bottom left) and 9.0\,m (bottom right) with PMT coverages ranging from 10\%-20\%. For each 5\,MeV step, the data points and shaded range correspond to the mean value and associated standard deviation respectively.}
    \label{fig:n9fit}
\end{figure}

The key variables used for antineutrino events of interest are the reconstructed time of the event vertex and the positron energy. In order to minimise the effect of scattered light within the detector and dark noise from the PMTs, only PMT hits from a narrow 9\,ns time window are used, 3\,ns before and 6\,ns after the light peak of a reconstructed event. This number of PMT hits, \nnine, is used as energy estimator. The relationship between the original energy of a particle and the reconstructed \nnine~is determined for each detector configuration by simulating positrons uniformly distributed within the inner detector volume with energies ranging from 0 to 80\,MeV, in steps of 5\,MeV. For each step, the mean \nnine~(energy scale) and observed width of the \nnine~distribution (energy resolution) are determined using a Gaussian fit. The mean \nnine-values were fitted using a second-order polynomial fit, $p_{0}x^2+p_1x+p_2$, to the input energy and resultant \nnine, with the linear part describing the lower energy behaviour ($E_{e}<15$\,MeV) and the quadratic part describing the higher energy section where saturation effects reduce the total number of PMT hits. This behaviour can be seen in figure~\ref{fig:n9fit} for all four tank configurations and PMT coverages ranging from 10\% to 20\%. Similarly, the width of the Gaussians was fitted to produce a parametric description of the energy resolution, using a polynomial of the form $p_{0}x^2+p_1x+p_2+p_3/x^{1/2}$.

As the event vertex reconstruction below 3\,MeV is poor, the prompt signal is required to deposit at least 3\,MeV measured energy to ensure correct reconstruction. Due to the high average antineutrino energy, as seen in figure~\ref{fig:modelspectra}, this has a minimal effect on supernova antineutrino rates. Similarly, a total energy of 3\,MeV is required for the neutron event. Events without coincident prompt (positron) and delayed (neutron) signals are rejected. In order to determine the overall detector efficiency for supernova bursts, each detector configuration has been simulated using a PMT coverage of 10\%, 15\% and 20\%. As an input spectrum, the first 500\,ms of a supernova burst using the Nakazato model were used, resulting in the total supernova event detection efficiencies listed in table~\ref{tab:efficiency}. This analysis is derived from reactor antineutrino-based analyses, which have shown a low background rate for double coincidence analysis on the order of $N_{bkg} < \mathcal{O}(10)$\,events per week. As a result, the overall background rate for the 500\,ms time window used in this analysis is negligible.

\begin{table}[ht]
\centering
    \begin{tabular}{lcccc}
    \hline
    & \multicolumn{4}{c}{Inner Detector Radius [m]} \\
    {PMT Coverage} & 5.7         & 6.7         & 7.7         & 9.0 \\
    \hline
    10\%            & 78.5\%   & 79.2\%   & 80.4\%   & 83.6\% \\
    15\%            & 82.2\%   & 83.4\%   & 83.6\%   & 85.1\% \\
    20\%            & 85.3\%   & 86.8\%   & 87.0\%   & 88.2\% \\
    \hline
    \end{tabular}
    \caption{Overview of total detection efficiencies of different detector configurations for IBD events during a supernova burst event, based on the Nakazato model.}
    \label{tab:efficiency}
\end{table}

\section{Detector response parametrisation}
\label{sec:responsemodel}

As supernova discrimination analysis requires a large number of supernova burst events and associated IBD events, it is not feasible to simulate the full response event-by-event for all detector configurations, due to computational requirements of the full photon simulation performed by RAT-PAC (up to 1\,s per event for high energy events). Instead, the detector response is approximated using the parametrised description of the energy scale and resolution in section~\ref{sec:simreco} to convert \textsc{sntools} output directly into the corresponding observable, \nnine. This allows the generation of a large amount of supernova events and the corresponding detector response without the need for the full RAT-PAC simulation.

\begin{figure}
    \centering
    \includegraphics[width=0.45\textwidth]{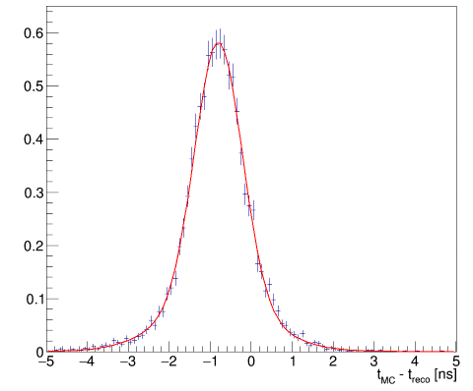}
    \caption{Time difference between input time \tmc\ and reconstructed time \treco\ for a detector with 9.0\,m inner detector radius with 20\% PMT coverage. The red curve represents a Gaussian fit to the distribution.}
    \label{fig:timedelta}
\end{figure}

The vertex timing difference between the generated time, \tmc, and reconstructed time, \treco, was also investigated. The result, shown in figure~\ref{fig:timedelta}, demonstrates that any differences are on the nanosecond scale and can be approximated by Gaussian smearing. Since the typical time scale between IBD events during a supernova burst with hundreds of events is on the order of $\mu$s to ms, this timing offset is expected to have a negligible impact on the analysis. The subsequent analysis has been performed with both \tmc\ and an approximated \treco\ with no discernible difference in the results.

\section{Statistical analysis}
\label{sec:stats}

In order to distinguish supernova models based on the observed spectrum, a comparison method between observed spectrum and the predicted underlying spectra is required. Using a modified Loredo-Lamb Approach~\cite{LoredoLamb89, HyperKDiscrimination}, each generated individual IBD event of the supernova burst and its corresponding time and energy value are compared against the expectation spectrum. This results in the following likelihood calculation for each full supernova burst and model:

\begin{equation}
    L=\mathrm{ln}\mathcal{L}=\sum_{i=1}^{N_{obs}}\mathrm{ln}N_{i}
\end{equation}

\noindent where $i$ runs over all detected IBD events,  $N_{obs}$, and $N_{i}$ is the number of events predicted by the model of choice in an infinitesimal bin centred on the time and energy values for IBD event~$i$. This likelihood is computed for each test model and the model with the largest likelihood is chosen as the identified model. If the identified model matches the model used to simulate the supernova burst in question, it is counted as correct identification, if the wrong model is identified, it is counted as misidentification. For each detector configuration, this procedure is performed repeatedly using 1,000 - 10,000 pseudoexperiments (PE) to determine the overall accuracy of the detector's ability to distinguish models. Each model has different identification and misidentification rates as some models are more easily confused with each other than others, leading to a range of possible discrimination accuracies for each detector configuration.

\section{Model discrimination results}
\label{sec:results}

By combining the detector efficiencies determined in section~\ref{sec:simreco} with the event rate predictions by \textsc{sntools}, the expected number of supernova IBD events at the benchmark distance of 10\,kpc is determined. Furthermore, as different models predict different antineutrino fluxes, this can be used to ascertain the distance at which the expectation value of $N_{obs}=100$ IBD events is predicted for each model, as shown in table~\ref{tab:distances}. These expected values assume the normal neutrino mass hierarchy.

\begin{table}[ht]
    \centering
    \begin{tabular}{lccc}
    \hline
    Model           & \shortstack{Interactions\\at 10\,kpc} & \shortstack{Detected IBD events\\at 10\,kpc} & \shortstack{Distance for\\100 IBD events [kpc]}\\
    \hline
    Nakazato        & 138.6                   & 122.2                          & 11.1 \\
    Vartanyan       & 359.3                   & 316.9                         & 17.8 \\
    Warren          & 290.3                   & 257.0                        & 16.0 \\
    \hline
    \end{tabular}
    \caption{The number of IBD interactions and expected number of detected IBD interactions at a supernova distance of 10\,kpc as well as the required distance for an expected number of $N_{obs}=100$ detected IBD events, based on the 9.0\,m ID radius (22\,m tank diameter), 20\% PMT coverage detector configuration.}
    \label{tab:distances}
\end{table}

\begin{figure}[ht]
    \centering
    \includegraphics[width=0.49\textwidth]{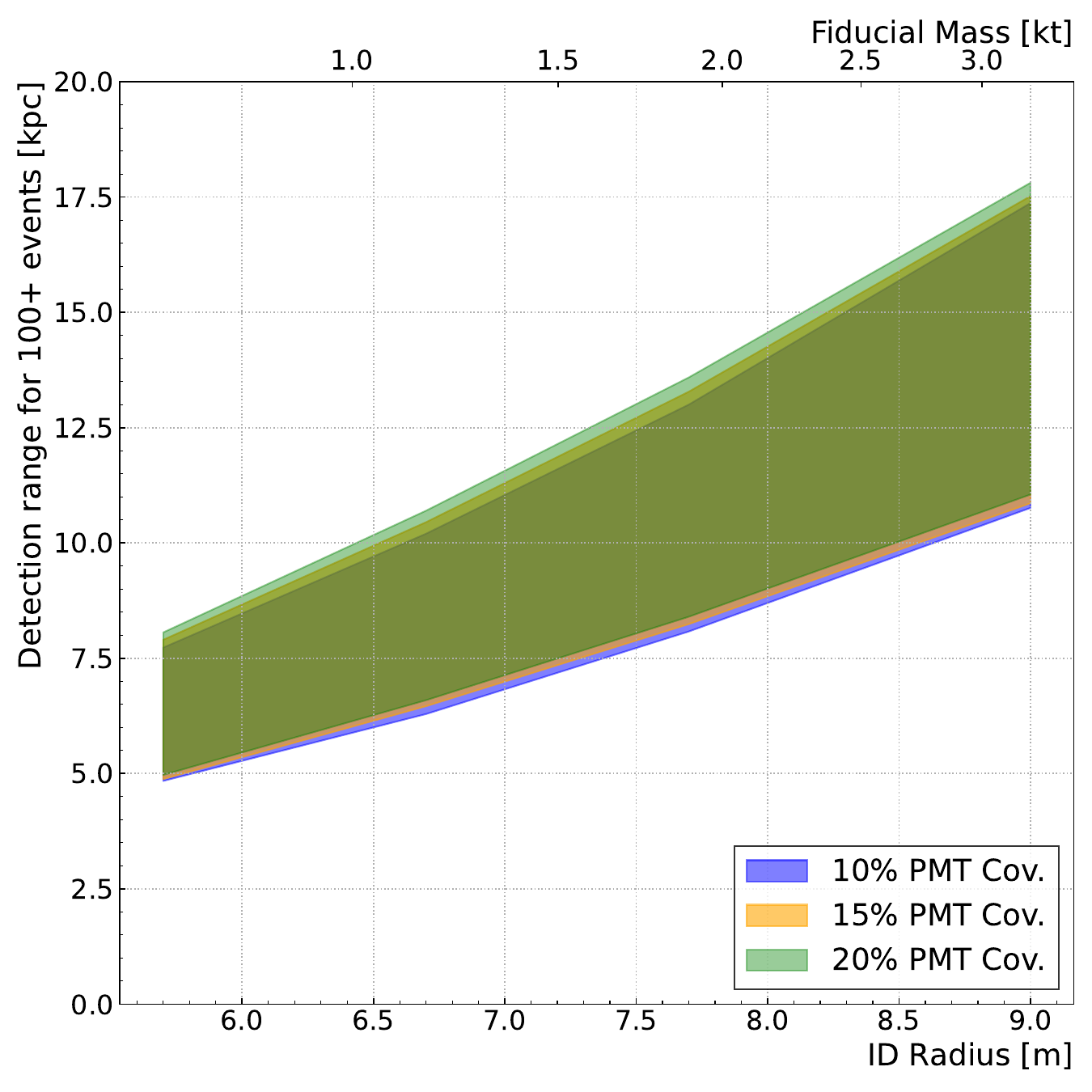}
    \includegraphics[width=0.49\textwidth]{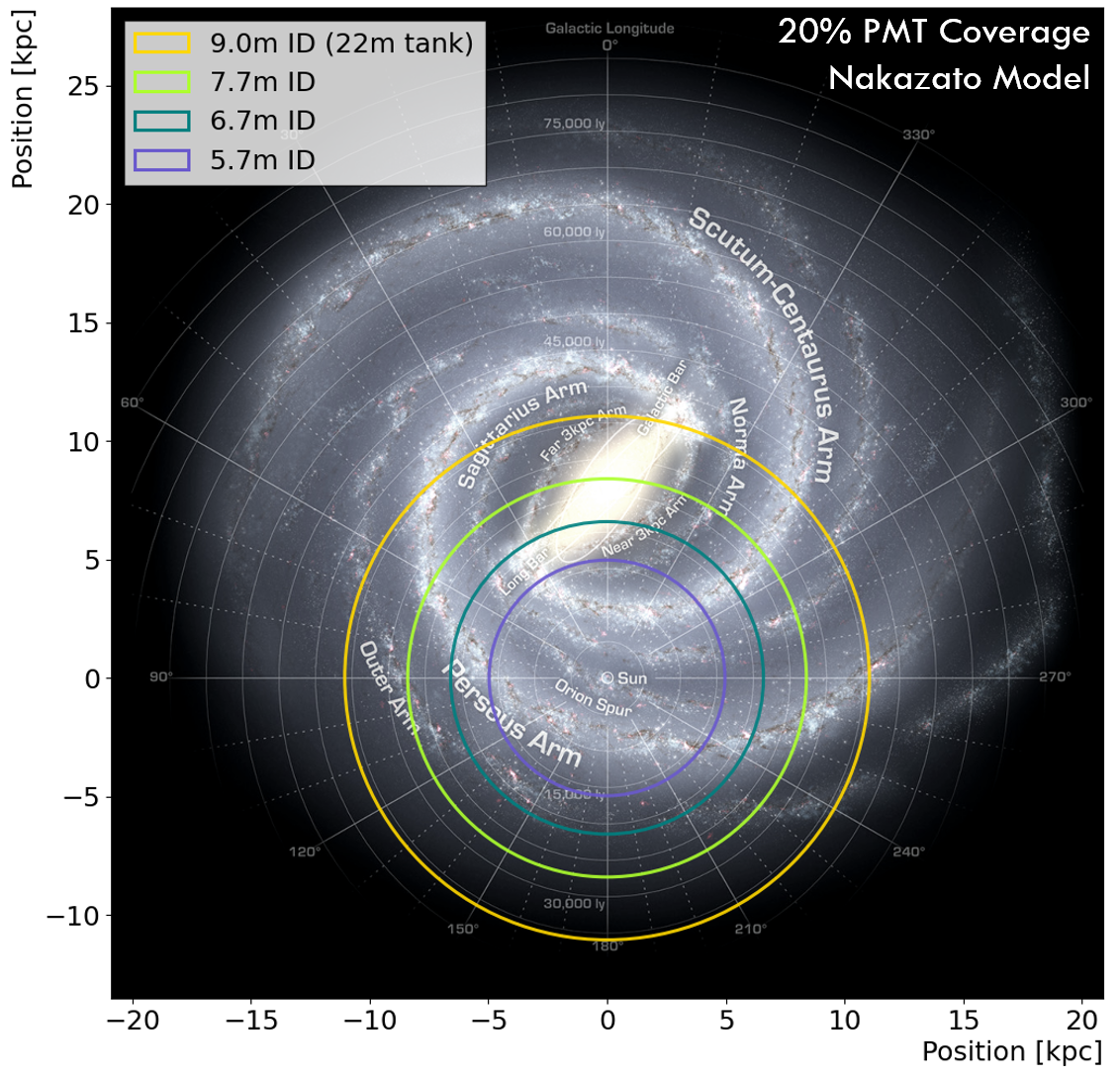}
    \caption{Maximum distance for a supernova to yield an expected number of 100 IBD events or more as a function of detector size (left) and effective range superimposed on the Milky Way, using the Nakazato model and a detector PMT coverage of 20\% (right). Galaxy image based on~\cite{GALAXY}.}
    \label{fig:galaxymap}
\end{figure}

\begin{figure}
    \centering
    \includegraphics[width=0.75\textwidth]{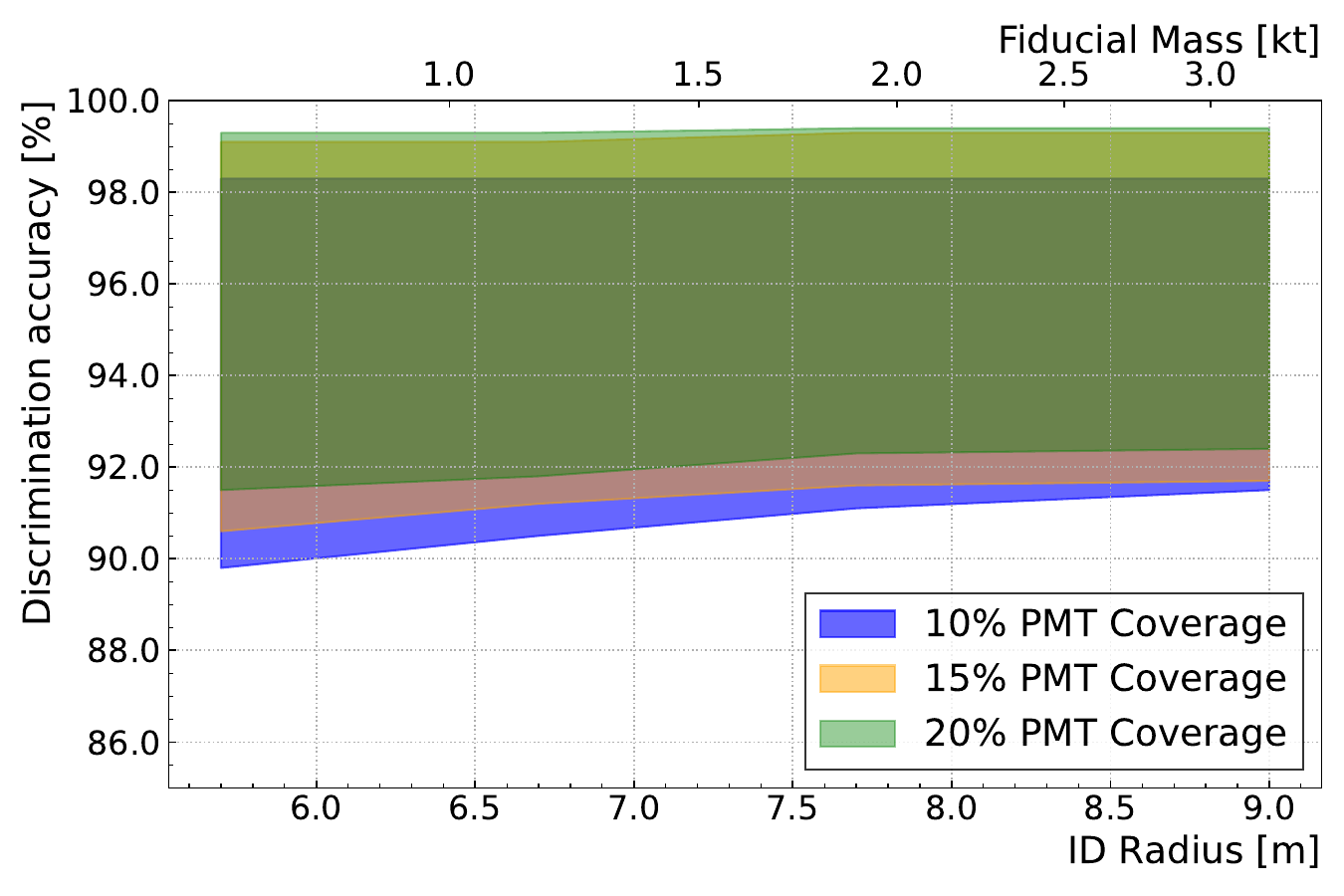}
    \caption{Model discrimination accuracy for different detector sizes and PMT coverages using a fixed event count of $N_{obs}=100$.}
    \label{fig:accu-by-100}
\end{figure}

By performing the statistical analysis outlined in section~\ref{sec:stats}, the model discrimination accuracy for a fixed number of events, $N_{obs}=100$, can be determined. This is done by generating 1,000 PE for each detector configuration and model, totalling 36,000 PEs, and performing the model discrimination analysis on each PE. The outcome is summarised in table~\ref{tab:100ev_comp} and figure~\ref{fig:accu-by-100}, showing that for 100 IBD events, the overall discrimination accuracy is consistently high ($>90$\%) and larger tank sizes have a negligible impact on the performance. Furthermore, an increase in PMT count yields minor benefits by increasing the minimum discrimination accuracy ($>91$\%), as the increased energy resolution afforded by the increased PMT coverage allows for better discrimination of hard-to-distinguish models.

\begin{table}[ht]
    \centering
    \begin{tabular}{lcccc}
    \hline
    & \multicolumn{4}{c}{Inner Detector Radius [m]} \\
    {PMT Coverage} & 5.7         & 6.7         & 7.7         & 9.0 \\
    \hline
    10\%            & 89.8 - 98.3\%   & 90.5 - 98.3\%   & 91.1 - 98.3\%   & 91.5 - 98.3\% \\
    15\%            & 90.6 - 99.1\%   & 91.2 - 99.1\%   & 91.6 - 99.3\%   & 91.7 - 99.3\% \\
    20\%            & 91.5 - 99.3\%   & 91.8 - 99.3\%   & 92.3 - 99.4\%   & 92.4 - 99.4\% \\
    \hline
    \end{tabular}
    \caption{Model discrimination accuracy for different detector sizes using $N_{obs}=100$.}
    \label{tab:100ev_comp}
\end{table}

These accuracy calculations can be combined with the distance calculations above to visualise the effective reach of the detector. By calculating the distance for each tank configuration with an expectation value of 100 IBD events, figure~\ref{fig:galaxymap} effectively displays the maximum distance for a tank configuration and supernova to perform at the determined discrimination accuracy. As the expected count rate of antineutrino events decreases as $1/r^{2}$, this also shows that the larger detector configurations are capable of observing multiple IBD events for any intragalactic supernova, making these configurations valuable as an early warning instrument.

\begin{figure}
    \centering
    \includegraphics[width=0.75\textwidth]{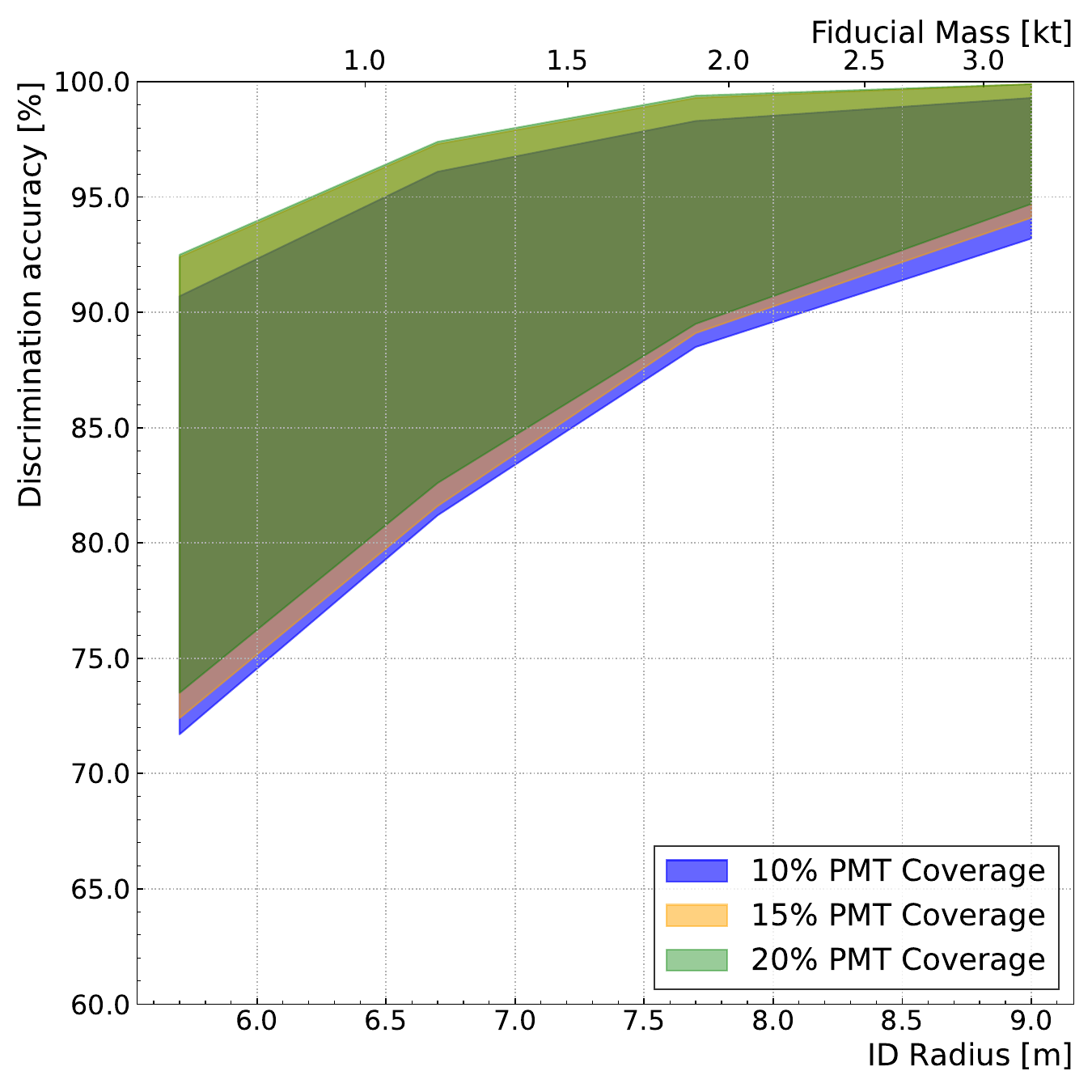}
    \caption{Model discrimination accuracy for different detector sizes and PMT coverages using a benchmark supernovae at 10\,kpc distance and event rates based on the Nakazato model.}
    \label{fig:accu-by-size}
\end{figure}

The strong dependence on the supernova distance,  $d_{SN}$, makes it evident that a comprehensive comparison requires a supernova at a fixed benchmark distance while varying the detector configuration. Since the varying flux from different models, seen in table~\ref{tab:distances}, would make the discrimination between the Warren model and the other two models trivial, the expected number of IBD events, $N_{obs}$, is calculated using the Nakazato model, acting as effective normalisation. The Nakazato model was chosen as the expected rate is the lowest of the three models, making it a conservative choice. The commonly used benchmark distance of $d_{SN}=10$\,kpc was chosen for easy comparison with other experiments~\cite{Snews, Al_Kharusi_2021}. This results in the number of events shown in table~\ref{tab:no_events}. Due to the lower number of IBD events at smaller tank sizes, the statistical fluctuation between PEs is higher, requiring 10,000 PEs per model and configuration to achieve converging accuracy results, for a total of 360,000 PEs. This model discrimination procedure results in the different discrimination accuracy shown in figure~\ref{fig:accu-by-size} and table~\ref{tab:distanceev_comp}.

\begin{table}
    \centering
    \begin{tabular}{lcccc}
    \hline
    & \multicolumn{4}{c}{Inner Detector Radius [m]} \\
    {PMT Coverage} & 5.7         & 6.7         & 7.7         & 9.0 \\
    \hline
    10\%            & 23    & 40    & 65    & 116 \\
    15\%            & 24    & 42    & 68    & 118 \\
    20\%            & 25    & 44    & 71    & 122 \\
    \hline
    \end{tabular}
    \caption{The number of events expected for different detector sizes using a fixed distance supernova event at $d_{SN} = 10$\,kpc according to the Nakazato model.}
    \label{tab:no_events}
\end{table}

\begin{table}[ht]
    \centering
    \begin{tabular}{lcccc}
    \hline
    & \multicolumn{4}{c}{Inner Detector Radius [m]} \\
    {PMT Coverage} & 5.7         & 6.7         & 7.7         & 9.0 \\
    \hline
    10\%            & 71.7 - 90.7\%   & 81.2 - 96.1\%   & 88.5 - 98.3\%   & 93.2 - 99.3\% \\
    15\%            & 72.4 - 92.4\%   & 81.6 - 97.3\%   & 89.1 - 99.3\%   & 94.1 - 99.9\% \\
    20\%            & 73.5 - 92.5\%   & 82.6 - 97.4\%   & 89.5 - 99.4\%   & 94.7 - 99.9\% \\
    \hline
    \end{tabular}
    \caption{Model discrimination accuracy for different detector sizes at a fixed distance of $d_{SN}=10$\,kpc.}
    \label{tab:distanceev_comp}
\end{table}

These results clearly show the large impact detector size has on the model discrimination performance of an antineutrino detector and the comparatively small effect of PMT coverage. This is due to the large increase in detector mass as the radius of a detector is increased and the use of timing information in addition to the energy information, reducing the usefulness of improved energy resolution, especially at supernova antineutrino energies. By performing this analysis using the selected detector configurations, it is also possible to interpolate between those configurations to determine the expected performance at a certain tank size.

\section{Estimation of early warning capabilities}
\label{sec:earlywarning}
Providing an early warning of an imminent supernova has been a frequently discussed capability of neutrino detectors, especially in the context of multi-messenger astronomy~\cite{Snews}. While this analysis has been developed to understand the supernova discrimination ability of a kilotonne-scale Gd-H$_2$O detector, the detector characterisation and analysis framework can also be used to estimate its capability to detect early warning signals. In this context, two distinct phases of a star's neutrino emission are of interest, the increased pre-supernova neutrino flux during the silicon burning phase of a star~\cite{ODRZYWOLEK2004303, odrzywolek2010neutrino} and the neutrinos produced during the core-collapse itself, as their arrival precedes the arrival of the optical signal.

Due to the similarities of the detector concept in geometry, fill medium and reconstruction to the Super-Kamiokande detector with Gd doping, a previous study~\cite{Machado_2022} has been used to estimate the IBD rate from silicon burning antineutrinos as $\sim$1.11-1.90 neutron events (0.32-0.56 with inverted neutrino mass hierarchy) per kiloton for Betelgeuse during the final eight hours before the core-collapse. Hence, $\sim$0.72-6.10 detectable events (0.21-1.80 with inverted neutrino mass hierarchy) are expected from Betelgeuse during these eight hours, depending on exact detector configuration. As this detector is intended to be located near reactors, producing antineutrinos in a similar energy range, the combined reactor antineutrino and background rate acts as effective background to the silicon burning signal and is expected to be $463.46 \pm 65.35$ events per kiloton per year ($\sim$1.30\ events per kiloton per day)~\cite{PhysRevApplied.19.034060}. This corresponds to $\sim$0.42\ background events per kiloton in eight hours, meaning in optimistic scenarios (normal neutrino mass hierarchy, stellar models), it could be possible to observe a silicon burning signal for relatively close stars, such as Betelgeuse, assuming the larger detector configurations and an optimised analysis.

During a supernova core-collapse, the total neutrino flux extends beyond the 500\,ms interval considered for the model discrimination. As the entirety of supernova burst can be used to provide early warning for optical telescopes, \textsc{sntools} has been used to estimate the total number of IBD events. The Nakazato model provides data for full 20\,s post core bounce and is therefore be used to determine the total number of expected IBD events in the detector. Analogous to the silicon burning estimate, the SNEWS parameters for Super-Kamiokande are used as a benchmark for the required detector performance~\cite{Snews}. Based on the expected $\sim$1.30\ effective background events per kiloton per day, a triggering burst is required to consist of $N_{obs} \geq 5$ detected events with a visible energy of $E_{e} \geq 7$\,MeV to minimise false alerts. Combining these parameters with the detector parameterisation performed in section~\ref{sec:simreco} results in distances ranging from 32.7 to 74.1\,kpc as shown in table~\ref{tab:earlywarning}. These results show that all detector configurations are capable of covering the entirety of the Milky Way, even assuming the conservative neutrino rates predicted by the Nakazato model. The largest detector configuration with a fiducial mass of 3.21\,kt can even provide a warning signal for supernovae in the Small Magellanic Cloud ($d_{SMC}\approx$63\,kpc).

\begin{table}[ht]
    \centering
    \begin{tabular}{cccc}
    \hline
    \shortstack{Inner Detector\\Radius [m]}           & \shortstack{Total interactions\\at 10\,kpc} & \shortstack{Total detected\\ IBD events at 10\,kpc} & \shortstack{Distance for\\5+ IBD events [kpc]}\\
    \hline
    5.7       & 66.2                    & 53.6                         & 32.7 \\
    6.7       & 118.1                   & 97.4                         & 44.1 \\
    7.7       & 192.4                   & 159.9                        & 56.6 \\
    9.0       & 326.8                   & 273.8                        & 74.0 \\
    \hline
    \end{tabular}
    \caption{Total number of IBD interactions from the full supernova burst (20\,s interval), expected number of detected IBD interactions at a supernova distance of 10\,kpc with $E_{e} \geq 7$\,MeV and the maximum distance for $N_{obs}\geq5$ detected IBD events, based on the 20\% PMT coverage detector configuration.}
    \label{tab:earlywarning}
\end{table}

\section{Summary and outlook}
\label{sec:summaryoutlook}
This study explores the supernova discrimination ability of a kilotonne-scale Gd-H$_2$O Cherenkov detector using the WATCHMAN detector concept. It demonstrates that the total fiducial volume and, by extension overall tank size, is the most important factor to its performance to observe and extract physics information from a core-collapse supernova.

This study shows that an application-driven antineutrino detector of sufficient size (18\,m or larger in diameter) in a low background environment is capable of making useful measurements at the commonly used benchmark distance of 10\,kpc, a distance within which approximately half of the supernova progenitors in our galaxy can be found. Additionally, these results show that the detector is capable of observing a significant number of events from a core-collapse supernova within the Milky Way and beyond and its potential to contribute to distributed observation using multiple neutrino telescopes and multi-messenger astronomy approaches, making it a valuable candidate for the inclusion in supernova early warning efforts. This demonstrates that, in addition to its original non-proliferation application, the WATCHMAN detector concept is a useful multipurpose platform for ongoing fundamental physics and science collaboration.

Furthermore, as this study shares its analysis and reconstruction tools with the reactor antineutrino-based tools, it highlights their versatility for use in liquid-based detectors but also means that a dedicated reconstruction optimised for supernova events may yield future improvement of these results.

\acknowledgments

The authors would like to thank M. Bergevin for the major contributions to the RAT-PAC software used for this analysis.

This work was supported by the Science and Technology Facilities Council (STFC).


\bibliography{sn-article}

\providecommand{\href}[2]{#2}\begingroup\raggedright\begin{thebibliography}{10}

\bibitem{burns2018remote}
J.~Burns, \emph{Remote detection of undeclared nuclear reactors using the {WATCHMAN} detector},  \href{https://arxiv.org/abs/1804.00655}{{\ttfamily 1804.00655}}.

\bibitem{osti_1544490}
A.~Bernstein, \emph{{Conceptual Design Overview of the Advanced Instrumentation Testbed (AIT) and the WATer CHerenkov Monitor of ANtineutrinos (WATCHMAN)}},  Tech. Rep. \href{https://www.osti.gov/biblio/1544490}{LLNL-TR-773250}, LLNL (2019), \href{https://doi.org/10.2172/1544490}{DOI}.

\bibitem{RevModPhys.92.011003}
A.~Bernstein, N.~Bowden, B.L.~Goldblum, P.~Huber, I.~Jovanovic and J.~Mattingly, \emph{Colloquium: Neutrino detectors as tools for nuclear security}, \href{https://doi.org/10.1103/RevModPhys.92.011003}{\emph{Rev. Mod. Phys.} {\bfseries 92} (2020) 011003}.

\bibitem{Adams_2013}
S.M.~Adams, C.S.~Kochanek, J.F.~Beacom, M.R.~Vagins and K.Z.~Stanek, \emph{Observing the next galactic supernova}, \href{https://doi.org/10.1088/0004-637x/778/2/164}{\emph{The Astrophysical Journal} {\bfseries 778} (2013) 164}.

\bibitem{JANKA200738}
H.-T.~Janka et~al., \emph{Theory of core-collapse supernovae}, \href{https://doi.org/https://doi.org/10.1016/j.physrep.2007.02.002}{\emph{Physics Reports} {\bfseries 442} (2007) 38}.

\bibitem{Vissani_2014}
F.~Vissani, \emph{Comparative analysis of {SN1987A} antineutrino fluence}, \href{https://doi.org/10.1088/0954-3899/42/1/013001}{\emph{Journal of Physics G: Nuclear and Particle Physics} {\bfseries 42} (2014) 013001}.

\bibitem{Snews}
A.~Habig and K.~Scholberg, \emph{The supernova early warning system}, \href{https://doi.org/https://doi.org/10.1038/s42254-020-0221-5}{\emph{Nature Reviews Physics} {\bfseries 2} (2020) 458}.

\bibitem{Al_Kharusi_2021}
S.A.~Kharusi, S.Y.~BenZvi, J.S.~Bobowski, W.~Bonivento, V.~Brdar, T.~Brunner et~al., \emph{{SNEWS} 2.0: a next-generation supernova early warning system for multi-messenger astronomy}, \href{https://doi.org/10.1088/1367-2630/abde33}{\emph{New Journal of Physics} {\bfseries 23} (2021) 031201}.

\bibitem{HyperKDiscrimination}
K.~Abe, P.~Adrich, H.~Aihara, R.~Akutsu, I.~Alekseev, A.~Ali et~al., \emph{Supernova model discrimination with {Hyper-Kamiokande}}, \href{https://doi.org/10.3847/1538-4357/abf7c4}{\emph{The Astrophysical Journal} {\bfseries 916} (2021) 15}.

\bibitem{VogelAndBeacom}
P.~Vogel and J.F.~Beacom, \emph{Angular distribution of neutron inverse beta decay, ${\overline{\ensuremath{\nu}}}_{e}+\stackrel{\ensuremath{\rightarrow}}{p}{e}^{+}+n$}, \href{https://doi.org/10.1103/PhysRevD.60.053003}{\emph{Phys. Rev. D} {\bfseries 60} (1999) 053003}.

\bibitem{STRUMIA200342}
A.~Strumia and F.~Vissani, \emph{Precise quasielastic neutrino/nucleon cross-section}, \href{https://doi.org/https://doi.org/10.1016/S0370-2693(03)00616-6}{\emph{Physics Letters B} {\bfseries 564} (2003) 42}.

\bibitem{2004gadzooks}
J.F.~Beacom and M.R.~Vagins, \emph{Antineutrino spectroscopy with large water Čerenkov detectors}, \href{https://doi.org/10.1103/physrevlett.93.171101}{\emph{Physical Review Letters} {\bfseries 93} (2004) }.

\bibitem{ROBINSON2003347}
M.~Robinson, V.~Kudryavtsev, R.~Luscher, E.~McMillan, P.~Lightfoot, N.~Spooner et~al., \emph{Measurements of muon flux at 1070m vertical depth in the {Boulby} underground laboratory}, \href{https://doi.org/https://doi.org/10.1016/S0168-9002(03)01973-9}{\emph{Nuclear Instruments and Methods in Physics Research Section A} {\bfseries 511} (2003) 347}.

\bibitem{Migenda2021}
J.~Migenda, S.~Cartwright, L.~Kneale, M.~Malek, Y.-J.~Schnellbach and O.~Stone, \emph{sntools: An event generator for supernova burst neutrinos}, \href{https://doi.org/10.21105/joss.02877}{\emph{Journal of Open Source Software} {\bfseries 6} (2021) 2877}.

\bibitem{Nakazato2013}
K.~Nakazato et~al., \emph{Supernova neutrino light curves and spectra for various progenitor stars: From core collapse to proto-neutron star cooling}, \href{https://doi.org/10.1088/0067-0049/205/1/2}{\emph{The Astrophysical Journal Supplement Series} {\bfseries 205} (2013) 2}.

\bibitem{Vartanyan2021}
A.~Burrows and D.~Vartanyan, \emph{Core-collapse supernova explosion theory}, \href{https://doi.org/10.1038/s41586-020-03059-w}{\emph{Nature} {\bfseries 589} (2021) 29–39}.

\bibitem{Baxter:2021xyq}
A.~Baxter et~al., \emph{{SNEWPY: A Data Pipeline from Supernova Simulations to Neutrino Signals}}, \href{https://doi.org/10.21105/joss.03772}{\emph{J. Open Source Softw.} {\bfseries 6} (2021) 03772}.

\bibitem{SNEWS:2021ewj}
{\scshape SNEWS} collaboration, \emph{{SNEWPY: A Data Pipeline from Supernova Simulations to Neutrino Signals}}, \href{https://doi.org/10.3847/1538-4357/ac350f}{\emph{Astrophys. J.} {\bfseries 925} (2022) 107} [\href{https://arxiv.org/abs/2109.08188}{{\ttfamily 2109.08188}}].

\bibitem{Warren2020}
M.L.~Warren et~al., \emph{Constraining properties of the next nearby core-collapse supernova with multimessenger signals}, \href{https://doi.org/10.3847/1538-4357/ab97b7}{\emph{The Astrophysical Journal} {\bfseries 898} (2020) 139}.

\bibitem{ratpac}
S.~Seibert, T.~Bolton, D.~Gastler, J.~Klein, H.~Lippincott, A.~Mastbaum et~al., \emph{{RAT-PAC (Reactor Analysis Tool, Plus Additional Codes)}},  2014.

\bibitem{AGOSTINELLI2003250}
S.~Agostinelli et~al., \emph{Geant4—a simulation toolkit}, \href{https://doi.org/https://doi.org/10.1016/S0168-9002(03)01368-8}{\emph{Nuclear Instruments and Methods in Physics Research Section A} {\bfseries 506} (2003) 250}.

\bibitem{ALLISON2016186}
J.~Allison, K.~Amako, J.~Apostolakis, P.~Arce, M.~Asai, T.~Aso et~al., \emph{Recent developments in {Geant4}}, \href{https://doi.org/https://doi.org/10.1016/j.nima.2016.06.125}{\emph{Nuclear Instruments and Methods in Physics Research Section A} {\bfseries 835} (2016) 186}.

\bibitem{fons_rademakers_2019_3457396}
F.~Rademakers et~al., \emph{root-project/root: Patch release of v6.18 series},  Sept., 2019.
\newblock 10.5281/zenodo.3457396.

\bibitem{PhysRevC.84.014306}
A.~Chyzh, B.~Baramsai, J.A.~Becker, F.~Be\ifmmode \check{c}\else \v{c}\fi{}v\'a\ifmmode~\check{r}\else \v{r}\fi{}, T.A.~Bredeweg, A.~Couture et~al., \emph{Measurement of the $^{157}\mathrm{Gd}$($n$,$\ensuremath{\gamma}$) reaction with the {DANCE} $\ensuremath{\gamma}$ calorimeter array}, \href{https://doi.org/10.1103/PhysRevC.84.014306}{\emph{Phys. Rev. C} {\bfseries 84} (2011) 014306}.

\bibitem{SMY2007118}
M.~Smy, \emph{Low energy challenges in {Super-Kamiokande-III}}, \href{https://doi.org/https://doi.org/10.1016/j.nuclphysbps.2007.02.065}{\emph{Nuclear Physics B - Proceedings Supplements} {\bfseries 168} (2007) 118}.

\bibitem{CARMINATI2015666}
G.~Carminati, \emph{The new wide-band solar neutrino trigger for {Super-Kamiokande}}, \href{https://doi.org/https://doi.org/10.1016/j.phpro.2014.12.068}{\emph{Physics Procedia} {\bfseries 61} (2015) 666}.

\bibitem{LoredoLamb89}
T.J.~Loredo and D.Q.~Lamb, \emph{{Neutrinos from SN 1987A}}, \href{https://doi.org/https://doi.org/10.1111/j.1749-6632.1989.tb50547.x}{\emph{Annals of the New York Academy of Sciences} {\bfseries 571} (1989) 601}.

\bibitem{GALAXY}
{NASA/JPL-Caltech/R. Hurt (SSC/Caltech)}, ``{The Milky Way Galaxy}.'' \url{https://solarsystem.nasa.gov/resources/285/the-milky-way-galaxy/}, 2017.

\bibitem{ODRZYWOLEK2004303}
A.~Odrzywolek, M.~Misiaszek and M.~Kutschera, \emph{Detection possibility of the pair-annihilation neutrinos from the neutrino-cooled pre-supernova star}, \href{https://doi.org/https://doi.org/10.1016/j.astropartphys.2004.02.002}{\emph{Astroparticle Physics} {\bfseries 21} (2004) 303}.

\bibitem{odrzywolek2010neutrino}
A.~Odrzywolek and A.~Heger, \emph{Neutrino signatures of dying massive stars: From main sequence to the neutron star.}, {\emph{Acta Physica Polonica B} {\bfseries 41} (2010) }.

\bibitem{Machado_2022}
L.N.~Machado, K.~Abe, Y.~Hayato, K.~Hiraide, K.~Ieki, M.~Ikeda et~al., \emph{Pre-supernova alert system for super-kamiokande}, \href{https://doi.org/10.3847/1538-4357/ac7f9c}{\emph{The Astrophysical Journal} {\bfseries 935} (2022) 40}.

\bibitem{PhysRevApplied.19.034060}
O.~Akindele, A.~Bernstein, M.~Bergevin, S.~Dazeley, F.~Sutanto, A.~Mullen et~al., \emph{Exclusion and verification of remote nuclear reactors with a 1-kiloton $\mathrm{Gd}$-doped water detector}, \href{https://doi.org/10.1103/PhysRevApplied.19.034060}{\emph{Phys. Rev. Appl.} {\bfseries 19} (2023) 034060}.

\end{thebibliography}\endgroup
\bibliographystyle{JHEP}
\end{document}